\documentclass[twocolumn]{aastex62}

\graphicspath{{./}{figures/}}

\shorttitle{the Multi-Messenger Matrix}
\shortauthors{Margalit \& Metzger}

\begin{document}

\title{The Multi-Messenger Matrix: the Future of Neutron Star Merger Constraints on the Nuclear Equation of State}

\email{benmargalit@berkeley.edu}

\author{Ben Margalit}
\altaffiliation{NASA Einstein Fellow}
\affil{Astronomy Department and Theoretical Astrophysics Center, University of California, Berkeley, Berkeley, CA 94720, USA}

\author{Brian D. Metzger}
\affil{Department of Physics and Columbia Astrophysics Laboratory, Columbia University, 538 West 120th St., New York, NY 10027}



\begin{abstract}

The electromagnetic (EM) signal of a binary neutron star (BNS) merger depends sensitively on the total binary mass, $M_{\rm tot}$, relative to various threshold masses set by the neutron star (NS) equation-of-state (EOS), parameterized through the NS maximum mass, $M_{\rm TOV}$, and characteristic radius, $R_{1.6}$.  EM observations of a BNS merger detected through its gravitational wave (GW) emission, which are of sufficient quality to ascertain the identity of the merger remnant, can therefore constrain the values of $M_{\rm TOV}$ and $R_{1.6}$, given the tight connection between $M_{\rm tot}$ and the well-measured chirp mass.  We elucidate the present and future landscape of EOS constraints from BNS mergers, introducing the ``Multi-Messenger Matrix", a mapping between GW and EM measurables that defines the ranges of event chirp masses which provide the most leverage on constraining the EOS.
By simulating a population of BNS mergers drawn from the Galactic double NS mass distribution we show that $\sim 10$ joint detections can constrain $M_{\rm TOV}$ and $R_{1.6}$ to several percent level where systematic uncertainties may become significant. Current EOS constraints imply that most mergers will produce supramassive or hypermassive remnants, a smaller minority (possibly zero) will undergo prompt-collapse, while at most only a few percent of events will form indefinitely stable NSs.
In support of the envisioned program, we advocate in favor of LIGO/Virgo releasing chirp mass estimates as early as possible to the scientific community, enabling observational resources to be allocated in the most efficient way to maximize the scientific gain from multi-messenger discoveries.

\end{abstract}

\keywords{keyword1 --- keyword2 --- keyword3 --- keyword4}


\section{Introduction and Methodology}

The discovery of gravitational waves (GW) from the binary neutron star (BNS) merger GW170817 \citep{LIGO+17DISCOVERY}, and commensurately of electromagnetic (EM) counterpart emission across the frequency spectrum \citep{LIGO+17CAPSTONE}, has initiated a new era of ``Multi-Messenger Astrophysics".  GW170817 was followed by a prompt burst of gamma-rays \citep{LIGO+17Fermi}; thermal radiation from the radioactive decay of heavy $r$-process nuclei (e.g.~\citealt{Drout+17,Kasen+17}); and non-thermal X-ray/radio emission (e.g.~\citealt{Margutti+18}), likely produced by an off-axis relativistic jet similar to those favored to produce the short-duration gamma-ray bursts (GRBs) observed on-axis at cosmological distances (e.g.~\citealt{Eichler+89,Berger14}).  Joint messenger observations of BNS mergers offer an unprecedented opportunity to explore the dynamical processes at work in these cataclysmic events as well as fundamental properties of neutron-rich matter near and above nuclear saturation density.  

GW detectors, such as LIGO/Virgo, accurately measure several properties of the inspiraling binary, particularly the chirp mass $\mathcal{M}_{\rm c}$.  They can also be used to measure or constrain the tidal deformability of the inspiraling stars prior to their disruption (e.g.~\citealt{Read+09,Raithel+18,Annala+18,Most+18,Fattoyev+18,DeSoumi+18,LIGO+18EOS}).  However, the current generation of detectors are far less sensitive to the post-merger signal and thus of the ultimate fate of the merger remnant, such as whether and when a black hole is formed \citep{LIGO+17REMNANT}.  Here, EM observations provide a complementary view.  For example, the lifetime of the neutron star (NS) remnant prior to black hole formation may be encoded in the luminosities and colors of the kilonova as the result of differences in the neutron abundance (and thus of the synthesis of lanthanide nuclei) in the ejecta \citep{Metzger&Fernandez14,Lippuner+17} and in the ejecta kinetic energy due to magnetar spin-down (\citealt{Metzger&Bower14,Horesh+16,Fong+16}).

\begin{figure}
\includegraphics[width=0.5\textwidth]{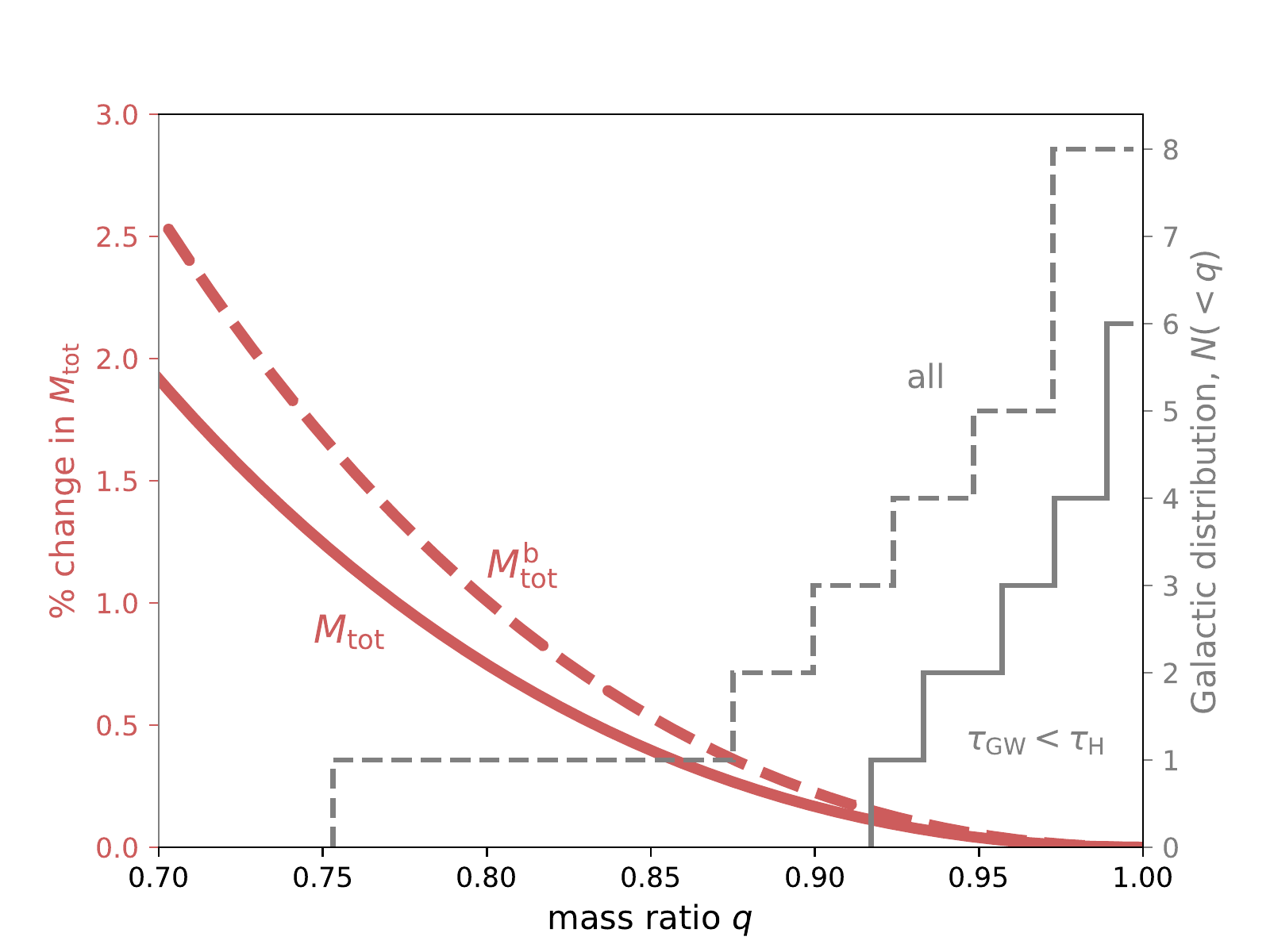}
\caption{Remnant mass of merging BNS system for a fixed chirp mass $\mathcal{M}_{\rm c}$, and as a function of the binary mass ratio $q$. The figure illustrates that both the gravitational and baryonic combined mass (solid and dashed red curves, respectively; eqs.~\ref{eq:Mrem_b},\ref{eq:m1m2}) depend weakly on $q$ within the range expected given the Galactic double NS distribution (grey cumulative histograms; solid --- systems of relevance, which will merge within a Hubble time).
A separate constraint that $m_2 \geq M_{\rm NS, min}$, where $M_{\rm NS, min} \approx 1.17 M_\odot$ is an estimate of the minimum NS mass thought to be able to form \citep{Suwa+18}, places a lower limit on $q$ (for fixed $\mathcal{M}_{\rm c}$) independent of assumptions about the NS spin or mass-ratio distributions. For GW170817 this yields $q \geq 0.73$.
}
\label{fig:q}
\end{figure}

The majority of the ejecta from GW170817, as inferred from modeling of the kilonova, likely originated in the post-merger phase from accretion disk outflows (e.g.~\citealt{Metzger+08,Fernandez&Metzger13,Perego+14,Just+15,Siegel&Metzger17,Fernandez+19}) rather than being the result of a dynamical ejection process.  This is important because key properties of the accretion disk, particularly its total mass and therefore its wind ejecta, are primarily functions of the total binary mass $M_{\rm tot}$ instead of the binary mass ratio $q$ (e.g.~\citealt{Radice+17,Coughlin+18b}, although see \citealt{Kiuchi+19}).  Although $q$ strongly affects the properties of the tidal dynamical ejecta, this component was likely a sub-dominant contributor in GW170817.  This is a fortunate circumstance because, although $q$ is poorly measured by GW observations, $M_{\rm tot}$ depends only weakly on $q$ for a precisely known value of $\mathcal{M}_{\rm c}$ (Fig.~\ref{fig:q}).   

The type of compact remnant created in the aftermath of a BNS coalescence event depends sensitively on its total mass, $M_{\rm tot}$, relative to various threshold masses.  These threshold masses depend on the NS equation of state (EOS), 
particularly those which determine key NS properties such as the 
maximal Tolman-Oppenheimer-Volkoff (TOV) mass $M_{\rm TOV}$ and the radius for a characteristic mass, e.g. $R_{1.6}$ for a $1.6M_{\odot}$ star.  The most massive binaries undergo `prompt' collapse to a black hole on the dynamical timescale (e.g.~\citealt{Shibata&Taniguchi06,Bauswein+13b}).  Slightly less massive binaries produce `hypermassive NS' (HMNS) remnants, which are temporarily stabilized from gravitational collapse by their rapid and differential rotation (e.g.~\citealt{Kastaun&Galeazzi15,Hanauske+17}).  Less massive binaries leave `supramassive NS' (SMNS) remnants, which remain supported against collapse even once in a state of rigid body rotation after their differential rotation has been removed (e.g.~by stresses due to strong internal magnetic fields).  Finally, the lowest mass binaries could in principle create remnants which are below the TOV mass and therefore indefinitely stable (e.g.~\citealt{Giacomazzo&Perna13}), even once their entire angular momentum has been removed (over potentially much longer timescales, e.g. by magnetic dipole spin-down).  

The expectation from numerical relativity that total mass is the dominant binary property which determines the outcome of a merger (a ``Vogt-Russell" theorem for BNS mergers) is one that a large sample of future mergers with both $M_{\rm tot}$ measurements and EM counterparts could test.  Likewise, with such a quantitative theoretical framework firmly in place, determination of whether an indefinitely stable NS, a SMNS, HMNS, or prompt-collapse occurs directly links EM and GW observation, permiting constraints on unknown NS EOS properties.  This method was utilized following the first BNS merger GW170817, where the measured $M_{\rm tot} = 2.74^{+0.04}_{-0.01}M_{\odot}$ \citep{LIGO+19PARAMS} along with the inference of a HMNS remnant from the kilonova colors and kinetic energy was used to place an upper limit of $M_{\rm TOV} \lesssim 2.17 M_\odot$ (\citealt{Margalit&Metzger17,Shibata+17}, and later corroborated by \citealt{Ruiz+18,Rezzolla+18}) as well as a lower limit on the NS radius $R_{1.6} \gtrsim 10.3-10.7$ km \citep{Bauswein+17,Radice+18}.\footnote{Prior to GW170817, a similar analysis was performed by \citet{Belczynski+08,Lawrence+15,Fryer+15}, who placed upper limits on $M_{\rm TOV}$ under the assumption that BH formation is a necessary requirement to explain the population of cosmological short GRBs if they indeed arise from merging NSs with the same mass distribution as known Galactic BNSs.} 

Here, we more systematically develop the multi-messenger approach to constraining the NS EOS.  A primary goal of this work is to elucidate the landscape of what can be learned about the EOS by a large sample of BNS mergers with well-characterized EM counterparts.  We further use this information to motivate search strategies and provide recommendations for information that LIGO/Virgo should promptly disseminate to the community of EM observers to maximize the scientific impact of its discoveries.  


\section{Multi-Messenger Matrix: predictions and learning opportunities}
\begin{figure*}
\center
\includegraphics[width=0.9\textwidth]{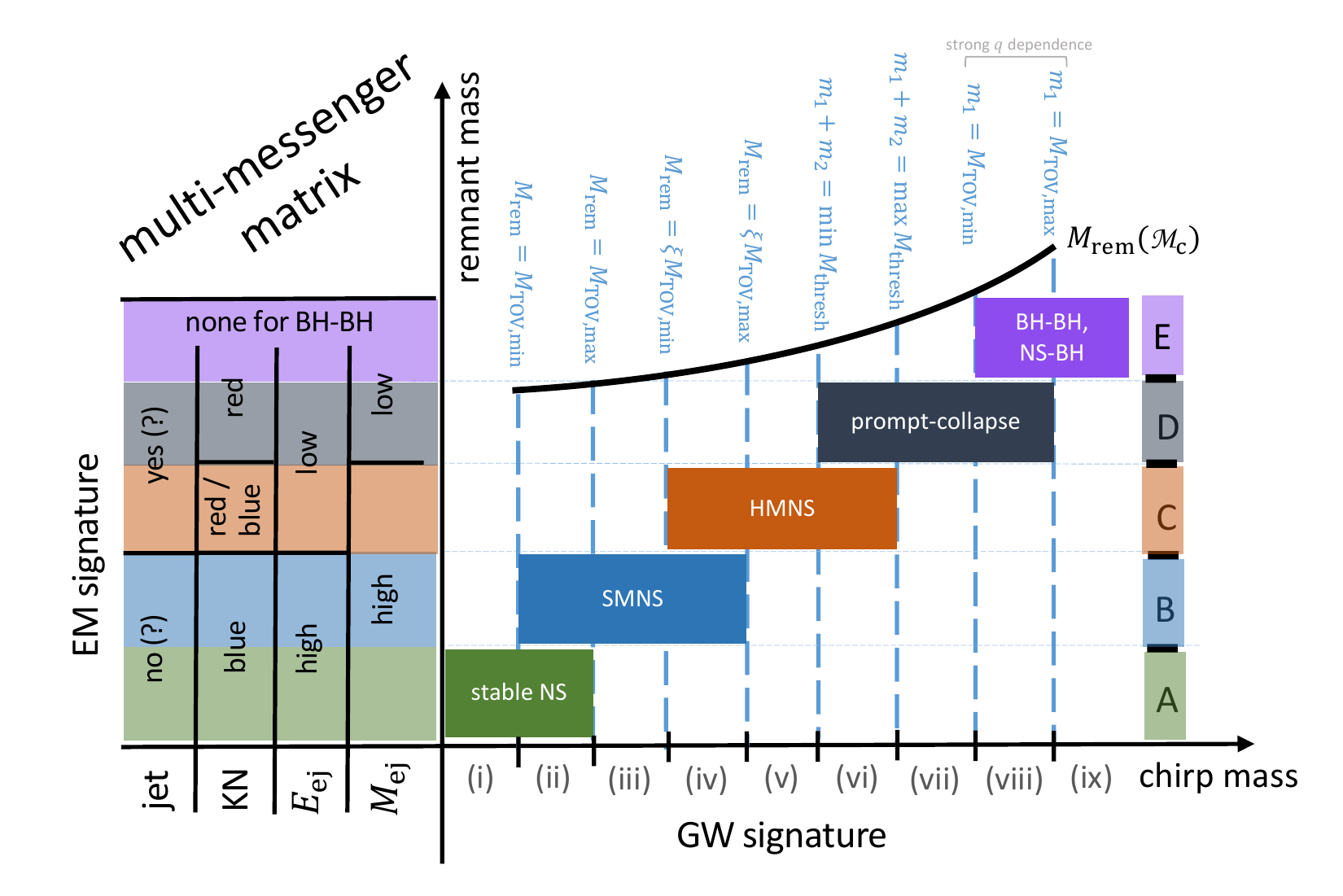}
\caption{``Multi-Messenger Matrix" relating the range of expected EM counterparts of binary neutron star mergers (vertical axis) to the GW-measured chirp mass (as a proxy for the binary mass; horizontal axis).  Depending on uncertain properties of the neutron star EOS ($M_{\rm TOV}$, $R_{1.6}$), the merger can result in one of four outcomes: indefinitely stable NS, long-lived SMNS, short-lived HMNS, or prompt BH formation.  Each outcome leads to qualitatively different predictions for the luminosity, color, and kinetic energy of the kilonova emission, as well as the possible presence or absence of an ultra-relativistic GRB jet.  Coupled with the GW-measured binary mass, an EM determination of the outcome of an individual merger event will either tighten constraints on the EOS properties (`learning opportunity') or serve as a consistency check on the model (`prediction') in those regions of parameter space where extant EOS constraints already determine the outcome.}
\label{fig:matrix}
\end{figure*}

Figure~\ref{fig:matrix} shows the current parameter space of GW and EM observables, delineated by the class of merger remnant. The total remnant mass (vertical axis), depicted by the black curve, is a function of the GW-measured chirp mass $\mathcal{M}_{\rm c}$ (horizontal axis), and only weakly dependent on the mass ratio $q$. 
The EOS sets mass cuts that determine the post-merger remnant for a given $\mathcal{M}_{\rm c}$.
Thus, different remnant types are separated along the vertical axis. Given that different remnants can imprint different observational signatures on the EM counterpart, this vertical axis can also be read as an ``EM signature'' axis.  

\begin{figure}
\includegraphics[width=0.5\textwidth]{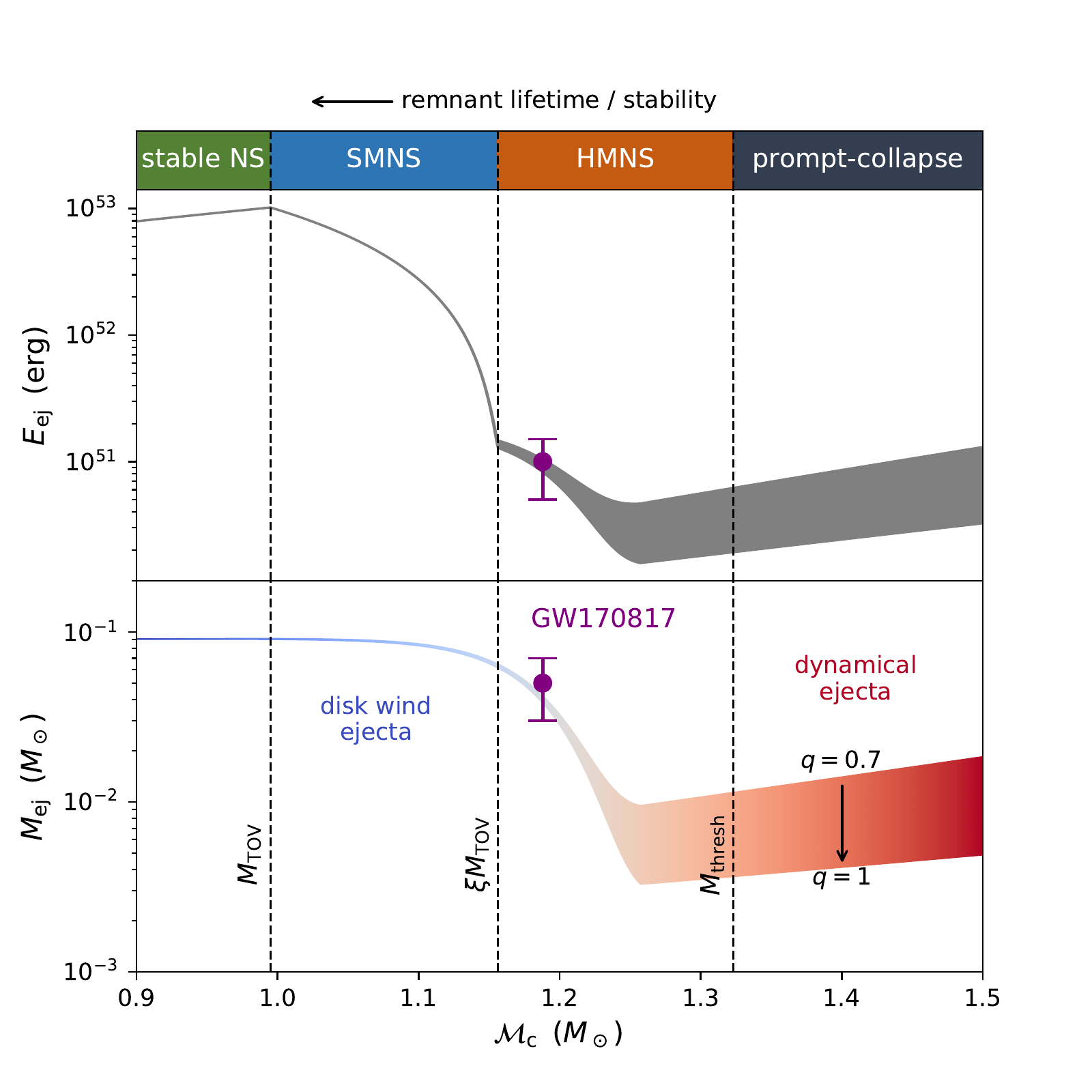}
\caption{Properties of the merger ejecta which affect the EM emission, such as the kilonova or synchrotron afterglow, as a function of the binary chirp mass.  Vertical dashed lines delineate the threshold masses for different merger remnants as marked, for an example EOS with $M_{\rm TOV} = 2.1M_{\odot}$ and $R_{\rm 1.6} = 12$ km. The top panel shows the ejecta kinetic energy, which we take to be the sum of the initial kinetic energy of the ejecta (estimated using fits to numerical relativity simulations; \citealt{Coughlin+18b,Coughlin+18a}) and, in the case of stable or SMNSs, the rotational energy which can be extracted from the remnant (\citealt{Margalit&Metzger17}).  The bottom panel shows the ejecta mass, both dynamical and disk wind ejecta, estimated as in \cite{Coughlin+18b}, where 50\% of the disk mass is assumed to be ejected at $v=0.15c$ (e.g. \citealt{Siegel&Metzger17}).  The finite width of the lines results from a range of binary mass ratio $q = 0.7-1$, to which the tidal dynamical ejecta is most sensitive.  The ejecta mass line is colored qualitatively according to the dominant color of the kilonova emission, which should become redder for more massive binaries (with shorter-lived remnants) due to their more neutron-rich ejecta (\citealt{Metzger&Fernandez14}).}
\label{fig:ejecta}
\end{figure}

The EM signatures listed in Fig.~\ref{fig:matrix} include the possible presence or absence of a relativistic GRB jet\footnote{The requirements for the successful launch of a GRB jet are still theoretically uncertain, and hence marked with question marks in Fig.~\ref{fig:matrix}. Magnetic field amplification in the HMNS might be necessary to produce a successful jet (e.g.~\citealt{Ruiz&Shapiro17}), which would imply that prompt-collapse/BH-NS mergers do not launch jets (however, see \citealt{Etienne+12,Paschalidis+15}). Baryon loading due to a neutrino-driven wind from the HMNS/SMNS remnant might also choke a putative jet prior to BH formation (e.g.~\citealt{Murguia-Berthier+14}). Conversely, the mass of the accretion disk orbiting the final BH following the collapse of a long-lived SMNS remnant is unlikely to be sufficient to power a GRB jet \citep{Margalit+15}.} (e.g.~\citealt{Murguia-Berthier+17}); the kilonova colors and the quantity of ejecta mass $M_{\rm ej}$, which depend on the NS remnant lifetime (e.g.~\citealt{Metzger&Fernandez14}); and the ejecta energy (e.g.~enhancement due to energy injection from a long-lived magnetar remnant; \citealt{Metzger&Piro14}).  Figure \ref{fig:ejecta} shows estimates of two measurable properties of the ejecta, particularly the total ejecta mass (measurable from the kilonova) and ejecta kinetic energy (measurable from the kilonova or synchrotron emission from interaction with the interstellar medium), as a function of the chirp mass for a range of $q$. The kilonova color is qualitatively shown in the bottom panel as well.

Uncertainties in our knowledge of the NS EOS translates into an ambiguity in the exact mass cut transitions between remnant types, or an overlap between possible merger outcomes (i.e. multiple possible EM observational signatures) given a measured GW chirp mass.
In the idealized approach adopted in this paper, this overlap is entirely a consequence of our uncertainty of the EOS.\footnote{Some intrinsic overlap between outcomes is to be expected due to additional, though likely second-order, effects such as: finite thermal effects on the EOS, binary mass ratio, etc.}
In other words, if the underlying EOS were precisely known, different measured chirp masses would have a one-to-one correspondence to differing EM signatures (different remnant types).
Regions in which no overlap between possible remnant types exists should therefore be interpreted as {\it predictions} --- if LIGO/Virgo measure a chirp mass in this region, our current understanding of the EOS implies only one possible merger outcome, giving a clear prediction for the type of EM counterpart expected to accompany such GW detection.
Conversely, chirp mass parameter space regions in which there is an overlap in the possible merger product should be seen as {\it learning opportunities} to constrain the EOS --- if an EM counterpart is observed in such an event and can be classified into one of the categories along the vertical axis, this would allow an inference of the merger outcome and break the ambiguity arising from our ignorance of the NS EOS.

\subsection{Merger Outcome Mass Cuts}

The cuts in chirp mass shown in Fig.~\ref{fig:matrix} are calculated from several criteria, mostly based on the remnant mass implied by a given chirp mass.
The remnant's baryonic mass depends on the well-measured chirp mass $\mathcal{M}_{\rm c}$, and weakly on the binary mass ratio $q$, but is also a function of the EOS through the translation between gravitational and baryonic masses:
\begin{equation}
M^{\rm b}_{\rm rem} = f(m_1) + f(m_2) - M_{\rm ej},
\label{eq:Mrem_b}
\end{equation}
where $M_{\rm ej}$ is the ejecta mass lost from the system,
\begin{equation}
m_1 = \mathcal{M}_{\rm c} q^{-3/5} (1+q)^{1/5} ,
~~~
m_2 = m_1 q ,
\label{eq:m1m2}
\end{equation}
are the masses of the two merging NSs, and
\begin{equation}
m^{\rm b} \equiv f(m) \approx m + 0.058 m^2 + 0.013 m^3
\label{eq:bindingenergy}
\end{equation}
is an approximate conversion between gravitational mass $m$ and baryonic mass $m^{\rm b}$ in units of solar mass accurate to within $10\%$ in binding energy ($m^{\rm b}-m$) for a winde range of piecewise polytropic EOSs (J.~Lattimer, private communication; \citealt{ZhaoLattimer18,Lattimer&Prakash16}).  Although, in detail, the function $f(m)$ is not universal and depends on the underlying EOS, we use this approximate expression in our current analysis for simplicity.
This is a reasonable approximation given that a $\pm 10\%$ deviation in binding energy induces only $<1\%$ variation in the chirp-mass thresholds derived below (see Tab.~\ref{tab:table}).

The maximal mass of a non-rotating NS (the TOV mass) is one of the poorly constrained properties of NSs that depends sensitively on the pressure at the highest densities achieved in the NS core.  Extant constraints on its value, $M_{\rm TOV, min} \lesssim M_{\rm TOV} \lesssim M_{\rm TOV, max}$ are nevertheless sufficient information for defining the different mass cuts illustrated in Fig.~\ref{fig:matrix}.

If the mass of the merger remnant is less than the minimum allowed TOV mass, $M_{\rm TOV, min}$, then the remnant is itself guaranteed to be indefinitely stable.  This criterion can be translated into a cut on chirp mass, 
$M_{\rm rem}^{\rm b}(\mathcal{M}_{\rm c}) < M_{\rm TOV, min}^{\rm b}$ (region (i) in Fig.~\ref{fig:matrix}).
For slightly larger chirp/remnant masses we have $M_{\rm TOV, min} < M_{\rm rem} < M_{\rm TOV, max}$, making it unclear whether $M_{\rm rem}$ is above or below the stable threshold mass, $M_{\rm TOV}$.  If EM observations can distinguish a stable remnant ($M_{\rm rem}<M_{\rm TOV}$) from one forming an unstable SMNS ($M_{\rm rem}>M_{\rm TOV}$), then a new upper (lower) limit on $M_{\rm TOV}$ is obtained.  Such a distinction could be made, for instance, based on the inferred kinetic energy of the merger ejecta (e.g.~\citealt{Metzger&Bower14}), or from evidence for black hole formation arising from an abrupt drop in the X-ray light curve (e.g.~\citealt{Rowlinson+10}).

For $M_{\rm rem} > M_{\rm TOV,max}$ a stable NS cannot form.
The transition between a SMNS and HMNS remnant is well approximated by $M_{\rm rem} = \xi M_{\rm TOV}$ with the constant of proportionality $\xi \simeq 1.18$ largely independent on the EOS \citep{Cook+94,Lasota+96}. Thus, for $M_{\rm rem} < \xi M_{\rm TOV,min}$ we do not expect a HMNS as the final pre-collapse merger remnant type. As a result, the region in which $M_{\rm TOV,max} < M_{\rm rem} < \xi M_{\rm TOV,min}$ is predicted to have only one possible merger outcome --- a SMNS (region (iii) in Fig.~\ref{fig:matrix}).

The remnants of mergers with slightly larger inferred masses satisfying $\xi M_{\rm TOV,min} < M_{\rm rem} < \xi M_{\rm TOV,max}$ could form either a SMNS or HMNS remnant (region (iv)).  Again, if a BNS merger detected in this mass range with a well characterized EM counterpart could disentangle these two possibilities, then this would provide a more stringent constraint on $M_{\rm TOV}$.  This was the strategy employed following the inference that GW170817 likely formed a HMNS remnant to constrain $M_{\rm TOV,max}$ (e.g.~\citealt{Margalit&Metzger17}).  This inference was made based on the quantity, kinetic energy, and composition of the kilonova ejecta (Fig.~\ref{fig:ejecta}). 

For a more massive merging NS system, the remnant may undergo a prompt-collapse to a BH, which will be accompanied by systematically less massive and more neutron-rich ejecta, resulting in a less luminous and redder kilonova than for the HMNS case.   The threshold for such collapse depends on the dense matter EOS parameterized through $M_{\rm TOV}$, but also on a compactness parameter $\propto M_{\rm TOV}/R_{\rm 1.6}$. 
Numerical and semi-analytic work have investigated this dependence \citep{Bauswein+13,BausweinStergioulas17} and found that the remnant will promptly collapse if 
\begin{equation}
    M_{\rm tot} > M_{\rm thresh} \approx \left( 2.38 - 3.606 \frac{G M_{\rm TOV}}{c^2 R_{1.6}} \right) M_{\rm TOV} ,
    \label{eq:Mthresh}
\end{equation}
where $R_{1.6}$ is the radius of a $1.6 M_\odot$ NS.
The threshold limit above which the merger undergoes a prompt-collapse is therefore an increasing function of both $M_{\rm TOV}$ and $R_{1.6}$.
A HMNS remnant can be ruled out for $M_{\rm tot}(\mathcal{M}_{\rm c}) > \max M_{\rm thresh}$ and similarly a prompt collapse is ruled out if $M_{\rm tot}(\mathcal{M}_{\rm c}) < \min M_{\rm thresh}$, where the maximum (minimum) is determined by the maximal (minimal) NS radius and TOV mass allowed based on extant constraints, or by the lowest (highest) chirp-mass merger observed to undergo prompt-collapse (form a HMNS).
The sharp drop off in the predicted ejecta mass approaching the prompt collapse threshold (Fig.~\ref{fig:ejecta}) should produce a kilonova signature readily distinguishable from the HMNS case.
This has been used to constrain $R_{1.6}$ following GW170817 \citep{Bauswein+17}, and can further be applied to future detections (\citealt{Bauswein+17}; see also Tab.~\ref{tab:table}).

Finally, under standard astrophysical formation scenarios (e.g. excluding primordial BHs), BHs are not expected to form with masses less than $M_{\rm TOV}$, and hence the region where $m_1 < M_{\rm TOV,min}$ is ruled out for BH-BH or BH-NS mergers.\footnote{Although a detectable EM counterparts are generally not expected for BH-BH mergers which take place in the interstellar medium, this could in principle change if the merger occurs in a gas-rich environment \citep[e.g.][]{Bartos+17,Stone+17,Khan+18}}

Table~\ref{tab:table} summarizes the ranges in chirp mass which delineate different predicted remnant types as well as the EOS constraints that could be obtained from distinguishing the EM counterpart for a GW event in each category.  This table is calculated based on extant constraints on the NS EOS following GW170817: $M_{\rm TOV,min} = 2.01M_{\odot}$ \citep{Antoniadis+13}; $M_{\rm TOV,max} = 2.17M_{\odot}$ \citep{Margalit&Metzger17}; $R_{1.6,\rm min} = 10.3$ km \citep{Bauswein+17}; $R_{1.6,\rm max} = 13.5$ km \citep{DeSoumi+18}, but can be updated following future GW/EM events.
Except for the two highest chirp-mass cuts related to the BH/NS mass transition, the chirp-mass boundaries are only weakly dependant on binary mass ratio and ejecta mass. This dependence is also given (to first order) in Tab.~\ref{tab:table}.

\begin{deluxetable*}{cccc}

\tablecaption{Merger remnant scenarios, based on EOS constraints available after GW170817 \label{tab:table}}

\tablehead{
\colhead{$\mathcal{M_{\rm c}}$$^{a}$ $(M_\odot)$} & \colhead{Remnant Type$^{b}$} & \colhead{EOS Constraint$^{c}$} & \colhead{Predicted Fraction$^{d}$}
}

\startdata
 & NS (i-A) & - & $\ll 1 \%$  \\
$0.948 \times \left[ 1 + 0.390 M_{\rm ej} - \delta(q)^{f} \right]$ & \cline{1-3}
 & NS (ii-A) & $M_{\rm TOV}$ lower bound $\geq 2.01 M_\odot$ & $3 \%$ \\
 & SMNS (ii-B) & $M_{\rm TOV}$ upper bound $\leq 2.17 M_\odot$ & \\
$1.032 \times \left[ 1 + 0.353 M_{\rm ej} - \delta(q) \right]$ & \cline{1-3}
 & SMNS (iii-B) & - &  $18 \%$ \\
$1.103 \times \left[ 1 + 0.325 M_{\rm ej} - \delta(q) \right]$ & \cline{1-3}
 & SMNS (iv-B) & $M_{\rm TOV}$ lower bound $\geq 2.01 M_\odot$ & $47 \%$ \\
 & HMNS (iv-C) & $M_{\rm TOV}$ upper bound $\leq 2.17 M_\odot$ & \\
$1.188$ & \cline{1-3}
 & HMNS (vi-C) & $R_{1.6}$ lower bound $\geq 10.3 \, {\rm km}$ & $32 \%$ \\
 & prompt-collapse (vi-D) & $R_{1.6}$ upper bound & \\
$1.433 \times \left[ 1 + 2^{-1/5} \delta(q) \right]^{-1}$ & \cline{1-3}
 & prompt-collapse (vii-D) & - & $\ll 1 \%$ \\
$1.750 \times q^{3/5} \left[ (1+q)/2 \right]^{-1/5}$ & \cline{1-3}
 & prompt-collapse (viii-D) & - & - \\
 & BH-BH, NS-BH (viii-E) & $M_{\rm TOV}$ upper bound $\leq 2.17 M_\odot$$^{e}$ & \\
$1.889 \times q^{3/5} \left[ (1+q)/2 \right]^{-1/5}$ & \cline{1-3}
  & BH-BH, NS-BH (ix-E) & - & - \\
\enddata
\tablecomments{$^{a}$Threshold chirp masses for different merger outcomes. Systematic uncertainty of $\pm10\%$ in the approximate binding energy relation (eq.~\ref{eq:bindingenergy}) induces $< 1\%$ variation in the first three $\mathcal{M}_{\rm c}$ cuts (the rest are unaffected); $^{b}$Type of compact remnant formed from the merger, labeled following the scheme from Fig.~\ref{fig:matrix}; $^{c}$EOS constraint obtained by ascertaining the type of merger remnant from EM observations; $^{d}$Calculated assuming a cosmological population of BNS mergers with a mass distribution following that of the known Galactic BNS population; 
$^{e}$Or evidence of a formation channel for BHs with $M<M_{\rm TOV}$;
$^{f}$Here $\delta(q) \equiv \left[ q^{-3/5}(1+q)^{6/5} - 2^{6/5} \right]/2$ is a small correction factor such that $\delta(q) \lesssim 0.02$ for $0.7 \leq q \leq 1$, and $M_{\rm ej}$ is the ejecta mass in units of $M_\odot$.
}

\end{deluxetable*}

\section{Populations and Future Constraints}

The previous section outlined the extant parameter space and illustrated the predicted outcomes for the next binary NS merger given our current knowledge and uncertainty regarding the NS EOS. We now address the implications of these results applied to a large population of merging binary NSs.

\subsection{Merger Outcome Statistics}

\begin{figure}
\includegraphics[width=0.5\textwidth]{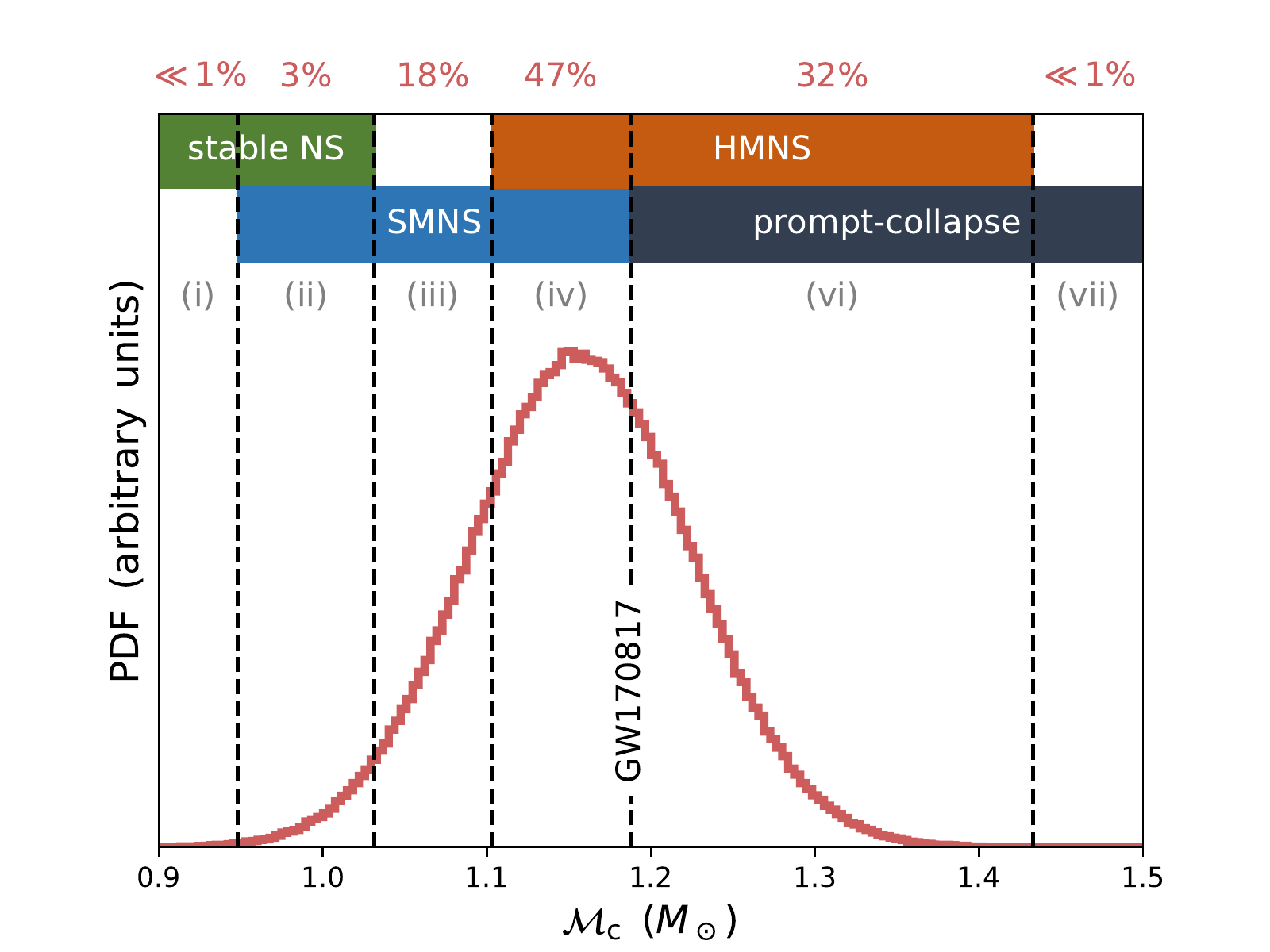}
\caption{Distribution of BNS merger chirp masses drawn from a NS population representative of Galactic double NSs \citep{Kiziltan+13}. The dashed vertical curves separate the $\mathcal{M}_{\rm c}$ parameter space based on the possible merger outcome(s) in each region, similar to Fig.~\ref{fig:matrix} (see Tab.~\ref{tab:table}). The fraction of mergers expected to occur in each region (the integral over the PDF within this region) is stated above the region in red. A significant fraction of mergers ($18\%-68\%$) should result in SMNS remnants, while only a small fraction $<3\%$ may produce indefinitely stable NSs. The fractions of mergers leading to HMNS remnants or prompt-collapse ranges from tens of percent to extremely infrequently.}
\label{fig:piechart}
\end{figure}

In order to model the population of extragalactic merging binary NS mergers, we assume the latter population is similar to that of known Galactic double NSs.  The inferred NS masses for GW170817 are consistent with being drawn from this population (e.g.~\citealt{LIGO+19PARAMS}; see also Fig.~\ref{fig:piechart}), which --- albeit for a sample size of $N=1$ --- supports this naive assumption.

The Galactic double NS population is narrowly peaked around $m_{1,2} \simeq 1.32 M_\odot$ with $q \approx 1$ \citep[e.g.][]{ZhaoLattimer18}. 
We draw NS masses from the Gaussian Galactic distribution found by \cite{Kiziltan+13}\footnote{Recent work suggests a possible bi-modality in the mass distribution \citep[e.g.][]{Farrow+19}, which however does not affect the distribution of $M_{\rm tot}$ of greater importance to this work.}, weighted by $\mathcal{M}_{\rm c}^{5/2}$. 
We include this chirp mass weighting for completeness, as it accounts for the GW detectors' bias of detecting higher-mass systems with larger GW-strain amplitudes (however, in practice we find negligible effect of this weighting on our final results).
Our analysis is similar to that of \cite{Piro+17}, with the important exception that \cite{Piro+17} examined specific example EOSs, thus fixing $M_{\rm TOV}$ and $R_{1.6}$. Here we instead calculate the statistics based on our current knowledge {\it and uncertainty} in the EOS. This gives realistic bounds on the fraction of different merger remnant types, and additionally allows us to estimate the number of mergers that will occur in regions of ambiguity in the remnant fate and thus can be used to further constrain the EOS.

Figure~\ref{fig:piechart} and and final column of Table \ref{tab:table} provide the expected fraction of mergers that will produce a given remnant.  We predict that the vast majority of mergers will result in either a HMNS or SMNS remnant ($\approx 18-68\%$ and $\approx 0-79\%$, respectively).  By contrast, the fraction that undergo prompt collapse could range from $\approx 0-32\%$.  Likewise, the fraction that end up as indefinitely stable remnants could range from $\approx 0-3\%$.
A potentially high fraction of SMNS remnants is consistent with findings based on analysis of short GRB X-ray plateaus \citep{Lasky+14,Gao+16} or temporally extended prompt X-ray emission \citep{Norris&Bonnell06,Perley+09}, assuming the latter are indicative of the presence of long-lived magnetar SMNS remnants (e.g.~\citealt{Metzger+08,Rowlinson+13,Zhang13}).  The rarity of indefinitely stable remnants, which may inject enormous rotational energy $\sim 10^{53}$ erg into the merger environment (Fig.~\ref{fig:ejecta}), is consistent with the tight limits available from radio transient surveys \citep{Metzger+15}.

\subsection{Future EOS Constraints}

\begin{figure}
\includegraphics[width=0.5\textwidth]{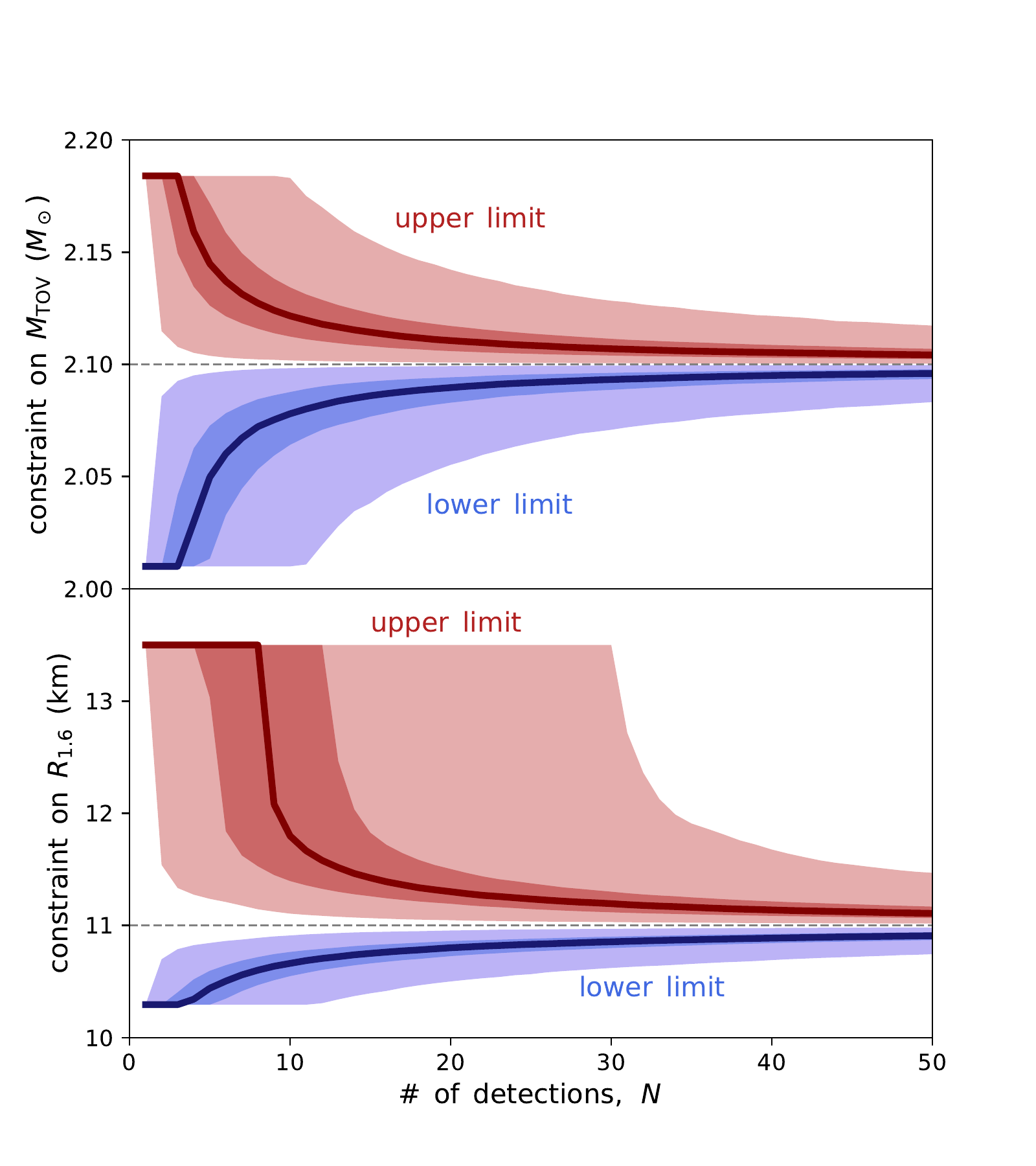}
\caption{As the number of detections $N$ with joint EM/GW determined outcomes increases, constraints on the NS EOS become tighter.  Here we show simulated constraints on the maximal NS mass, $M_{\rm TOV}$ (upper panel), and the radius of a $1.6 M_\odot$ NS, $R_{1.6}$ (bottom panel) as a function of $N$. Horizontal dashed curves show the assumed underlying real values of these parameters, here $M_{\rm TOV} = 2.1 M_\odot$ and $R_{1.6} = 11 \, {\rm km}$, while the red (blue) regions above (below) these curves depict the upper (lower) bounds on the parameters. Solid curves show the median constraint among the ensemble of simulated detections, while dark (light) shaded regions show the $1 \sigma$ ($2 \sigma$) variations around the median. }
\label{fig:Ndet}
\end{figure}

We now consider what improvement could be made in EOS constraints with a sample of future joint messenger detections.  Figure~\ref{fig:Ndet} shows the expected constraints on $M_{\rm TOV}$ and $R_{1.6}$ as a function of the number of binary NS merger detections with sufficiently well observed EM counterparts to accurately ascertain the remnant outcome.  
For this calculation we have assumed a particular EOS as being the ``correct" one, for which $M_{\rm TOV} = 2.1 M_\odot$ and $R_{1.6}=11$ km (horizontal dashed-grey curves in Fig.~\ref{fig:Ndet}).  We then draw a sample of $N-1$ binary NS masses from the Galactic binary distribution, as described in the previous subsection.  
This simulates a set of $N$ successive NS merger detections (the first of which is always taken to be GW170817). Given the chirp mass of the merging system and assuming that the observed EM counterpart is able to break the degeneracy between different possible merger remnants (see Figs.~\ref{fig:matrix},\ref{fig:ejecta}), we use the methods of \cite{Margalit&Metzger17,Bauswein+17} to update the constraints on $M_{\rm TOV}$ and $R_{1.6}$ following each detection. We then repeat this process for a newly drawn set of $N-1$ binary masses, creating a large ensemble of such detection sets.
The median upper (lower) bounds on $M_{\rm TOV}$ and $R_{1.6}$ obtained following $N$ detections are then shown by the dark red (blue) curves, with $68\%$ and $95\%$ percentiles around this median depicted by dark and light shadowed regions, respectively (where the median and percentiles are calculated over the ensemble).

We find that after $N \gtrsim 10$ mergers, one could in principle constrain the value of $M_{\rm TOV}$ and $R_{\rm 1.6}$ to within a few percent, close to the level where systematic uncertainties will set in (see discussion below).  The convergence on $R_{\rm 1.6}$ is sensitive to the true NS radius (e.g. significantly slower for $R_{1.6}=12$ km) because it is only constrained by the prompt collapse condition (\citealt{Bauswein+17}), and thus depends on whether the true $M_{\rm thresh}$ (eq.~\ref{eq:Mthresh}) is low enough to generate a large population of prompt collapse events.  

In Fig.~\ref{fig:Ndet} we have accounted only for EOS constraints obtained using the BNS merger multi-messenger approach described above (e.g. the $R_{1.6}$ constraint does not account for potential tidal-deformability limits placed by future BNS mergers). Additional NS observations can and should be folded in to even further constrain the EOS.
For example, well measured masses of massive NSs \citep{Cromartie+19,Antoniadis+13,Demorest+10}, experimental laboratory data, NS radius measurements, and the multi-messenger BNS merger constraints described in this work can be combined using the statistical framework described by \cite{Miller+19}, providing more stringent constraints than any single method used alone.


\section{Discussion}

Following the ground-breaking discovery, follow-up, and multi-messenger detection of the first binary NS merger GW170817 a flurry of theoretical works have used this watershed event to place constraints on fundamental physics, including the NS EOS.  Several of the methods and important results in this field came from analyzing the multi-messenger nature of this event and not, as previously envisioned, solely through GW waveform analysis.  These newly applicable methods have illustrated that important scientific results can be obtained from even a single well observed event such as GW170817. This is in contrast to the post-merger or inspiral GW constraints on the EOS, which are typically thought to require multiple events to build-up significant signal-to-noise and place meaningful constraints on the EOS (e.g.~\citealt{Chatziioannou+17,Torres-Rivas+19}).

Given the success of the multi-messenger approach for GW170817, now is an opportune time to explore the path ahead for this budding field.  The ongoing LIGO/Virgo O3 run may detect between $\sim 1-10$ additional BNS mergers, while such discoveries could occur as frequently as once per week following the planned LIGO A+ upgrade.  We have here summarized the current methodology for placing constraints on the NS EOS via such multi-messenger observations and laid out the post-GW170817 parameter space, highlighting regions of particular importance for EOS constraints, as well as providing explicit falsifiable predictions of our methods. 

Our envisioned program relies on extensive EM follow-up of GW-detected mergers, particularly those with chirp masses indicating they will provide the most leverage on the EOS if the outcome can be constrained (`learning opportunity'). Given the finite number of observing hours which can be dedicated to EM searches of LIGO/Virgo localization regions, and the large number of near future GW detections expected, it is inevitable that not all mergers will be followed-up with the most sensitive wide-field optical/near-IR facilities.
It is therefore crucial to identify as early as possible, and before a follow-up campaign is launched or aborted, the possible scientific impact of obtaining multi-messenger observations for each event.

An under-appreciated inference from GW170817 is that ejecta from the disk wind, rather than dynamically ejected material, is likely the most important source of the observed kilonova emission (Fig.~\ref{fig:ejecta}).  The properties of the kilonova emission is therefore likely to be more sensitive to the total binary mass (a proxy for stability of the remnant) than to the binary mass ratio $q$.  This is important because, while $q$ is poorly constrained by GW data, $M_{\rm tot}$ is tightly related to the well-measured chirp mass (Fig.~\ref{fig:q}).  As shown in Table \ref{tab:table}, if the GW-measured chirp mass is indeed the key indicator of the expected EM counterpart, then certain ranges of chirp mass values are more important than others in obtaining new and meaningful constraints on the NS EOS through the multi-messenger analysis.
We therefore strongly advocate in favor of releasing chirp mass estimates as early as possible to the scientific community, ideally already with the initial LIGO/Virgo triggers, so that the observational community will be able to allocate its resources in the most efficient way to maximize the scientific gain from multi-messenger observations.

Focusing on an idealized scenario, in which the binary mass is measured to infinite precision and the EM counterparts provide an accurate discriminator between different merger remnants, we find that the radius and maximum mass of NSs could be constrained to within a few percent after $\gtrsim 10$ events (Fig.~\ref{fig:Ndet}).  However, a number of systematic uncertainties will enter at the percent level, if not earlier, which limit the precision of the constraints.  First, although the chirp mass is generally measured to effectively infinite precision, percent-level uncertainty in the translation of $\mathcal{M}_{\rm c}$ to $M_{\rm tot}$ results because of the much larger uncertainty in $q$ (Fig.~\ref{fig:q}). 
Determining $M_{\rm rem}$ accurately also requires subtracting off $M_{\rm ej}$ (which is at best constrained from the kilonova emission to a factor of $\lesssim 2$; e.g.~\citealt{Wu+19}), and accounting for the NS binding energy (eqs.~\ref{eq:Mrem_b},\ref{eq:bindingenergy}) which implicitly depends on the true EOS.
Furthermore, the intrinsic chirp mass is degenerate with the redshift of the event; therefore as merger events begin to probe cosmological-scale volumes, uncertainties in the cosmological distance scale will ultimately propagate to those in the EOS.

The largest source of uncertainty in our outlined procedure likely come from our incomplete understanding of the merger physics.  We thus note the obvious need for further theoretical investigations into the late-time evolution of NS merger remnants.  These efforts should be two-fold --- focusing on both the post-merger dynamics and remnant formation (this has implications for the precise mass cuts separating different remnant classes) and an enhanced characterization of the EM signatures that may identify each remnant type (including a better understanding of natural sources of variability and possible contaminants).

Theoretical maps like Figure \ref{fig:ejecta} are based on simplified assumptions that do not account for some physical effects which could introduce quantitative uncertainties in the EOS constraints. These include thermal effects on the equation of state over the neutrino cooling timescale $\sim 1 \, {\rm s}$ (e.g.~\citealt{Kaplan+14}), and the precise way in which angular momentum is redistributed throughout the merger remnant.  For instance, 
\cite{Fujibayashi+18}
and \citet{Radice+18b} include the effects of an artificial $\alpha$-viscosity into GR hydro simulations of the post-merger evolution to mimic physical angular momentum transport, finding the presence of additional ejecta sources not included in the usual ``dynamical + disk wind" dichotomy.
In a recent study, \citet{Kiuchi+19} show that mass ratio can affect the prompt-collapse threshold and disk mass estimates, potentially motivating a revised examination of the prompt-collapse criteria.

The ``Vogt-Russell" theorem for BNS mergers which underpins the strategy advanced here is motivated by numerical relativity simulations, but must itself be tested with a large ensemble of events.  This highlights the importance of characterizing the EM counterparts of not just mergers in the mass range which tighten EOS constraints, but those in mass ranges for which the merger outcome is ``predicted".  For instance, our approach is calibrated on the results of numerical simulations that largely use a standard hadronic equation of state, which ignores the possibility of a first-order phase transition at high densities or temperatures to deconfined quarks (e.g.~\citealt{Paschalidis+18,Han+18,Drago+18,Most+19,Bauswein+19}).  Deviations in an ensemble of mergers from the predictions of a ``well-behaved" EOS might provide evidence for such a phase transition.  

Finally, we note that the multi-messenger methods described here rely solely on a {\it qualitative} categorization of which remnant type was produced by the merger. Even stronger EOS constraints can in principle be obtained by quantitatively fitting inferred kilonova properties to numerical relativity simulations \citep{Radice+17,Radice&Dai18,Coughlin+18a,Coughlin+18b,Lazzati&Perna19}. Such methods are somewhat more susceptible to uncertainties in the kilonova modelling and numerical relativity simulations, but are a promising way forward.

On the opposite end, GW-waveform analysis alone can provide an extremely clean model-independent probe of the NS radius however this method requires very high signal-to-noise events, especially at $\sim$kHz frequencies. Events as loud as GW170817 should be rare, implying that such constraints may not improve quickly. In contrast, the multi-messenger methods discussed above require only information about the chirp mass, which should be well measured even for lower signal-to-noise events, although much will hinge on the success rate of detecting and characterizing the EM counterparts to the requisite fidelity.
Dedicated follow-up programs on large survey instruments, such as the Large Synoptic Survey Telescope, could play a major role in the success of this endeavor (\citealt{Cowperthwaite+19}).

\section*{Acknowledgements}
We thank Jim Lattimer, Andreas Bauswein, and Kenta Kiuchi for helpful comments on the manuscript.
BM is supported by NASA through the NASA Hubble Fellowship grant \#HST-HF2-51412.001-A awarded by the Space Telescope Science Institute, which is operated by the Association of Universities for Research in Astronomy, Inc., for NASA, under contract NAS5-26555.
BDM is supported in part by NASA through the Astrophysics Theory Program (grant number NNX16AB30G).
The final stages of this work benefited from conversations at KITP, supported in part by the National Science Foundation under Grant No. NSF PHY-1748958.


\end{document}